\documentstyle[12pt]{article}

\newcommand {\nn}    {\nonumber}
\newcommand {\vs}[1]  { \vspace*{#1 cm} }

\newcounter{eq}
\newcounter{sc}

\newcommand {\NP}   {Nucl.Phys.}
\newcommand {\PL}   {Phys.Lett.}

\newcommand {\PRL}   {Phys.Rev.Lett.}

\newcommand {\JHEP}  {JHEP}

\def\overleftrightarrow#1{\vbox{\ialign{##\crcr
 $\leftrightarrow$\crcr\noalign{\kern-1pt\nointerlineskip}
 $\hfil\displaystyle{#1}\hfil$\crcr}}}

\newlength{\minitwocolumn}
\begin{document}


\begin{flushright}
EDO-EP-31\\
August, 2000\\
\end{flushright}
\vspace{30pt}

\pagestyle{empty}
\baselineskip15pt

\begin{center}
{\large\bf Bosonic Fields in the String-like Defect Model

 \vskip 1mm
}

\vspace{20mm}

Ichiro Oda
          \footnote{
          E-mail address:\ ioda@edogawa-u.ac.jp
                  }
\\
\vspace{10mm}
          Edogawa University,
          474 Komaki, Nagareyama City, Chiba 270-0198, JAPAN \\

\end{center}


\vspace{15mm}
\begin{abstract}
We study localization of bosonic bulk fields on a string-like
defect with codimension 2 in a general space-time dimension 
in detail. We show that in cases of spin 0 scalar and
spin 1 vector fields there are an infinite number of massless
Kaluza-Klein (KK) states which are degenerate with respect to the
radial quantum number, but only the massless zero mode state among 
them is coupled to fermion on the string-like defect. 
It is also commented on interesting extensions of the
model at hand to various directions such as 'little' superstring
theory, conformal field theory and a supersymmetric construction.

\vspace{15mm}

\end{abstract}

\newpage
\pagestyle{plain}
\pagenumbering{arabic}


\rm
\section{Introduction}

In the theories where our four dimensional world is a 3-brane
embedded in a higher dimensional space \cite{Rubakov, Akama, Visser,
Arkani-Hamed, Randall2}, the conventional Kaluza-Klein scenario 
would be modified drastically.
In most works on the Kaluza-Klein compactification
thus far, a higher dimensional manifold is assumed to be composed 
of as a direct product of a non-compact four dimensional Minkowski 
space-time and a compact internal manifold  
with the size of the compact space being set by the Planck scale.
However, in this approach it seems to be quite difficult to stabilize 
the size of all the internal dimensions around the Planck scale via some 
non-perturbative effects. This problem should be solved in the
brane world \footnote{In a supersymmetric model, flat directions 
could appear so that the stability problem of moduli seems at first
glance to be not so important as in a non-supersymmetric model. 
But in this case, we need the fine-tuning of the parameters.}.

In recent years, an alternative scenario of the compactification 
has been put forward \cite{Randall2}. This new idea is based on  
the possibility that our world is a 3-brane embedded in
a higher dimensional space-time with non-factorizable warped
geometry. In this scenario, we are free from the moduli 
stabilization problem in the sense that the internal
manifold is noncompact and does not need to be compactified to
the Planck scale any more, which is one of reasons 
why this new compactification scenario has attracted so much attention. 
An important ingredient of this scenario is that  
all the matter fields are thought of as confined to the a 3-brane, 
whereas gravity is free to propagate in the extra dimensions.
Such localization of matters would be indeed possible in D-brane 
theory \cite{Polchinski} and M-theory \cite{Horava}, 
but at present it is far from complete to realize the Rundall-Sundrum 
model \cite{Randall2} within the framework of superstring theory. 
Thus, it is worthwhile to explore whether such localization is 
also possible in the local field theory.
 
In fact, the localization mechanism has been recently investigated in 
$AdS_5$ space \cite{Rizzo, Pomarol, Grossman, Bajc, Chang}.  
In particular, it is shown that spin 0 field is localized
on a brane with positive tension which also localizes the graviton
\cite{Bajc}, while spin 1 field is not localized neither on a brane
with positive tension nor on a brane with negative tension
\cite{Pomarol, Bajc}. Moreover, it is shown that spin 1/2 and 3/2 
fields are localized not on a brane with positive tension but on a 
brane with negative tension \cite{Grossman, Bajc}. Thus, in order to 
fulfill the localization of Standard Model particles on a brane 
with positive tension, it seems that some additional interactions 
except gravity must be also introduced in the bulk. 

More recently, the possibility of extending the Randall-Sundrum
domain wall model to higher dimensional topological objects was
explored \cite{Chodos, Cohen, Gregory, Csaki, Vilenkin, 
Gherghetta, Minic, Chaichian, Chodos2, Oda, Gherghetta2}.
In particular, we find that Einstein's equations admit a string-like
defect with codimension 2 in addition to a domain wall with codimension
1 \footnote{In this terminology, topological
defects with codimension 3 and 4, respectively, would be called
a monopole-like defect and an instanton-like defect.}.
In particular, the existence of the string-like defect makes 
it possible to think of a 3-brane in the six dimensional 
anti-de Sitter space.

In this string-like defect model, the localization of bulk fields has
been also investigated. In Ref. \cite{Gherghetta}, it is shown that
spin 2 graviton is localized on the 3-brane and the corrections to 
Newton's law are more supressed than in the domain wall model. 
Afterwards, the present author has explored the localization of 
various spin fields on the string-like defect in a general dimension 
and obtained the following facts \cite{Oda} : spin 0, 1 and 2 bosonic fields 
are localized on a string-like defect with the exponentially decreasing 
warp factor, whereas spin 1/2 and 3/2 fermionic fields are 
localized on a defect with the exponentially increasing warp factor. 
These results for the localization of various spin fields
coincide with the corresponding ones \cite{Bajc}
in the Randall-Sundrum model \cite{Randall2} and many brane 
model \cite{Oda1, Oda2} except spin 1 vector field.
It is of interest that there is no localized vector field
on the brane in the domain wall model \footnote{See Ref. \cite{Dubovsky} 
for an interesting possibility of electric charge non-conservation 
in brane world where a higher-dimensional generalization
of the Randall-Sundrum model is used in order to localize
gauge fields on a brane.}, while vector
field can be localized on the defect in the string-like
model. This phenomenon can be briefly explained as follows:
In the Randall-Sundrum model, we can see that the overall
coefficient in front of gauge field action is divergent so that 
we do not have a normalizable zero mode of the bulk gauge field. 
On the other hand, in our string-like
model, we have an additional warped factor coming from part of the angular 
variable in the background metric in addition to the conventional
warped factor. Combined with these two warped factors, the coefficient
in front of the action becomes finite so the zero mode of the bulk gauge
field is 
normalizable and is consequently localized on the string-like defect.

One aim of the present paper is to investigate this interesting
property of bulk bosonic fields in the string-like defect model in
more detail. The case of spin 2 graviton field has been already
examined in Ref. \cite{Gherghetta, Gherghetta2}, 
so we will concentrate on
the study of spin 0 scalar and spin 1 vector fields. 
We will show that there are an infinite number of massless
Kaluza-Klein (KK) modes which are degenerate with respect to the
radial quantum number, but only one massless field among them is
coupled to fermion on the string-like defect.  
Moreover, the KK excitations of gauge field have vanishing coupling to 
spin 1/2 fermion on the defect, so gauge field can exist 
in the bulk without meeting  any phenomenological constraints 
on the model, which should be contrasted with the Randall-Sundrum 
domain wall model where the strong coupling of the KK excitations 
of gauge field to the brane fermion gave rise to a potential 
internal inconsistency within the theory \cite{Rizzo, Pomarol}.    

This paper is organized as follows. In the next section, we 
review a string-like defect solution with codimension 2. In
Section 3, the Kaluza-Klein decomposition of scalar field
is studied in a background obtained in Section 2. Then, in
Section 4, the procedure used in Section 3 is applied to the
case of gauge field. 
The final section is devoted to discussions.

\section{A string-like defect}

Let us start with a brief review of a string-like defect solution
to Einstein's equations with sources to fix our
notations and conventions  \cite{Oda}. 
We consider Einstein's equations with a bulk cosmological
constant $\Lambda$ and an energy-momentum tensor $T_{MN}$
in general $D$ dimensions:
\begin{eqnarray}
R_{MN} - \frac{1}{2} g_{MN} R 
= - \Lambda g_{MN}  + \kappa_D^2 T_{MN},
\label{2}
\end{eqnarray}
where $\kappa_D$ denotes the $D$-dimensional gravitational
constant.
Throughout this article we follow the standard 
conventions and notations of the textbook of Misner, Thorne and 
Wheeler \cite{Misner}. 

Let us adopt the following metric ansatz:
\begin{eqnarray}
ds^2 &=& g_{MN} dx^M dx^N  \nn\\
&=& g_{\mu\nu} dx^\mu dx^\nu + \tilde{g}_{ab} dx^a dx^b  \nn\\
&=& e^{-A(r)} \hat{g}_{\mu\nu} dx^\mu dx^\nu + dr^2 
+ e^{-B(r)} d \Omega_{n-1}^2,
\label{4}
\end{eqnarray}
where $M, N, ...$ denote $D$-dimensional space-time indices, 
$\mu, \nu, ...$ do $p$-dimensional brane ones, and $a, b, ...$
do $n$-dimensional extra spatial ones, so the equality $D=p+n$
holds. (We assume $p \ge 4$.) And d$\Omega_{n-1}^2$
stands for the metric on a unit $(n-1)$-sphere, which is 
concretely expressed in terms of the angular variables $\theta_i$ as
\begin{eqnarray}
d \Omega_{n-1}^2 = d\theta_2^2 + \sin^2 \theta_2 d\theta_3^2 
+ \sin^2 \theta_2 \sin^2 \theta_3 d\theta_4^2 + \cdots
+ \prod_{i=2}^{n-1} \sin^2 \theta_i d\theta_n^2.
\label{5}
\end{eqnarray}

Moreover, we shall take the ansatz for the energy-momentum tensor
respecting the spherical symmetry:
\begin{eqnarray}
T^\mu_\nu &=& \delta^\mu_\nu t_o(r),  \nn\\
T^r_r &=& t_r(r),  \nn\\
T^{\theta_2}_{\theta_2} &=& T^{\theta_3}_{\theta_3} = \cdots
= T^{\theta_n}_{\theta_n} = t_\theta(r),
\label{6}
\end{eqnarray}
where $t_i(i=o, r, \theta)$ are functions of only the radial
coordinate $r$. 

With these ansatzs, after a straightforward calculation,
Einstein's equations (\ref{2}) reduce to
\begin{eqnarray}
e^A \hat{R} - \frac{p(n-1)}{2} A' B' - \frac{p(p-1)}{4} (A')^2
- \frac{(n-1)(n-2)}{4} (B')^2 \nn\\
+ (n-1)(n-2) e^B - 2\Lambda + 2 \kappa_D^2 t_r = 0,
\label{7}
\end{eqnarray}
\begin{eqnarray}
e^A \hat{R} + (n-2) B'' - \frac{p(n-2)}{2} A' B' 
- \frac{(n-1)(n-2)}{4} (B')^2  \nn\\
+ (n-2)(n-3) e^B + p A'' -  \frac{p(p+1)}{4} (A')^2 - 2\Lambda 
+ 2 \kappa_D^2 t_\theta = 0,
\label{8}
\end{eqnarray}
\begin{eqnarray}
\frac{p-2}{p} e^A \hat{R} + (p-1)(A'' - \frac{n-1}{2} A' B')
- \frac{p(p-1)}{4} (A')^2 \nn\\
+ (n-1) [B'' - \frac{n}{4} (B')^2 + (n-2) e^B ]  
- 2\Lambda + 2 \kappa_D^2 t_o = 0,
\label{9}
\end{eqnarray}
where the prime denotes the differentiation with respect to $r$,
and $\hat{R}$ is the scalar curvature associated with the
brane metric $\hat{g}_{\mu\nu}$.
Here we define the cosmological constant on the $(p-1)$-brane, 
$\Lambda_p$, by the equation
\begin{eqnarray}
\hat{R}_{\mu\nu} - \frac{1}{2} \hat{g}_{\mu\nu} \hat{R} 
= - \Lambda_p \hat{g}_{\mu\nu}.
\label{10}
\end{eqnarray}
In addition, the conservation law for the energy-momentum tensor,
$\nabla^M T_{MN} = 0$ takes the form
\begin{eqnarray}
t'_r = \frac{p}{2} A' (t_r - t_o) + \frac{n-1}{2} B' (t_r - t_\theta).
\label{11}
\end{eqnarray}

Our purpose is to find a string-like defect solution, that is,
$n=2$, with a warp factor $A(r) = c r$ ($c$ is a positive constant)
to the above equations. (The case of $n=1$ corresponds to
a domain wall solution.) The necessity of this exponentially
decreasing warp factor is to bind gravity to the p-brane.
{}For generality, we consider
a general space-time dimension $D$ and a general brane dimension
$p$ with $D=p+2$, but the physical interest, of course,
lies in the case of six space-time dimensions ($D=6$) and a 
3-brane ($p=4$). 
In the case of $n=2$, under the ansatz $A(r) = c r$, 
Einstein equations (\ref{7}), (\ref{8}), (\ref{9}) are of the form
\begin{eqnarray}
e^{cr} \hat{R} - \frac{p}{2} c B' - \frac{p(p-1)}{4} c^2
- 2\Lambda + 2 \kappa_D^2 t_r = 0,
\label{12}
\end{eqnarray}
\begin{eqnarray}
e^{cr} \hat{R}  -  \frac{p(p+1)}{4} c^2 - 2\Lambda 
+ 2 \kappa_D^2 t_\theta = 0,
\label{13}
\end{eqnarray}
\begin{eqnarray}
\frac{p-2}{p} e^{cr} \hat{R} -  \frac{p-1}{2} c B'
- \frac{p(p-1)}{4} c^2 + B'' - \frac{1}{2} (B')^2   
- 2\Lambda + 2 \kappa_D^2 t_o = 0,
\label{14}
\end{eqnarray}
and the conservation law takes the form
\begin{eqnarray}
t'_r = \frac{p}{2} c (t_r - t_o) + \frac{1}{2} B' (t_r - t_\theta).
\label{15}
\end{eqnarray}

{}From these equations, general solutions can be found as follows:
\begin{eqnarray}
ds^2 = e^{-cr} \hat{g}_{\mu\nu} dx^\mu dx^\nu + dr^2
+ e^{-B(r)} d\theta^2,
\label{16}
\end{eqnarray}
where 
\begin{eqnarray}
B(r) = cr + \frac{4}{pc} \kappa_D^2 \int^r dr (t_r - t_\theta),
\label{17}
\end{eqnarray}
\begin{eqnarray}
c^2 &=& \frac{1}{p(p+1)}(-8 \Lambda + 8 \kappa_D^2 \alpha), \nn\\
\hat{R} &=& \frac{2p}{p-2} \Lambda_p = -2 \kappa_D^2 \beta.
\label{18}
\end{eqnarray}
Here $t_\theta$ must take a definite form, which is given by
\begin{eqnarray}
t_\theta = \beta e^{cr} + \alpha,
\label{19}
\end{eqnarray}
with $\alpha$ and $\beta$ being some constants. Moreover, in
order to guarantee 
the positivity of $c^2$, $\alpha$ should satisfy an inequality
$-8 \Lambda + 8 \kappa_D^2 \alpha > 0$.

Two types of special solution deserve more scrutiny. 
A specific solution is the one without sources
$(t_i = 0)$. Then we get a special solution which was found for
a $\it{local}$ string in Ref. [18], and for a $\it{global}$ 
string in Ref. [15]:
\begin{eqnarray}
ds^2 = e^{-cr} \hat{g}_{\mu\nu} dx^\mu dx^\nu + dr^2
+ R_0^2 e^{-cr} d\theta^2,
\label{20}
\end{eqnarray}
with $R_0$ being a length scale which we take to be of order unit. 
Here the positive constant $c$, the brane 
scalar curvature and the brane cosmological constant are respectively
given by
\begin{eqnarray}
c^2 &=& \frac{-8 \Lambda}{p(p+1)}, \nn\\
\hat{R} &=& \frac{2p}{p-2} \Lambda_p = 0.
\label{21}
\end{eqnarray}
In this case, as in the corresponding domain wall solution, the bulk 
geometry is the anti-de Sitter space, and the brane geometry is 
Ricci-flat with vanishing cosmological constant. 
It has been recently found that this special solution corresponds
to a $\it{local}$ defect in the sense that the energy-momentum tensor
is strictly vanishing outside the string core \cite{Gherghetta,
Gherghetta2}

Another specific solution occurs when we have the spontaneous 
symmetry breakdown $t_r = -t_\theta$ \cite{Vilenkin}:
\begin{eqnarray}
ds^2 = e^{-cr} \hat{g}_{\mu\nu} dx^\mu dx^\nu + dr^2
+ R_0^2 e^{-c_1 r} d\theta^2,
\label{22}
\end{eqnarray}
where
\begin{eqnarray}
c^2 &=& \frac{1}{p(p+1)}(-8 \Lambda + 8 \kappa_D^2 t_\theta) > 0, \nn\\
c_1 &=& c - \frac{8}{pc} \kappa_D^2 t_\theta, \nn\\
\hat{R} &=& \frac{2p}{p-2} \Lambda_p = 0.
\label{23}
\end{eqnarray}
Notice that this solution is more general than the previous one
(\ref{20}) since this solution reduces to (\ref{20}) when $t_\theta
= 0$.
In Ref. \cite{Gherghetta2}, the solution (\ref{22}) was called
a $\it{global}$ defect since there appears a hedgehog type 
configuration outside the string core.

To close this section, let us comment on an interesting
$\it{global}$ defect recently found in a general dimension
in Ref. \cite{Gherghetta2}.
To gain the $\it{global}$ topological defect, the antisymmetric
tensor field with rank $n-2$ is added to the Einstein-Hilbert 
action with a cosmological constant. Then the energy-momentum
tensor associated with the $(n-2)$-form field in the bulk has
the property
\begin{eqnarray}
t_0 = t_r = - t_\theta.
\label{23-2}
\end{eqnarray}
The ansatz taken in Ref. \cite{Gherghetta2} is 
\begin{eqnarray}
A(r) = cr, \ B(r) = constant.
\label{23-3}
\end{eqnarray}
With this ansatz (\ref{23-3}), it is easy to see that Einstein's 
equations (\ref{7}), (\ref{8}), (\ref{9}) and the conservation 
law (\ref{11}) require important equations
\begin{eqnarray}
t_0 = t_r = constant, t_\theta = constant,
\label{23-4}
\end{eqnarray}
in addition to the other inessential equations for the present 
consideration. 
These conditions (\ref{23-4})
is more general than (\ref{23-2}), so if an energy-momentum tensor
satisfies (\ref{23-4}), Einstein's equations with such energy-momentum
tensor would admit the $\it{global}$ topological defect with the background 
metric (\ref{23-3}) as a solution in a general space-time dimension.
Finally, note that this new $\it{global}$ defect has the same property
as the domain wall with respect to the localization of various bulk
fields.

\section{Kaluza-Klein decomposition of scalar field}

In previous paper, it was shown that spin 0, 1, and 2 bosonic
fields are localized on the p-brane defect with the exponentially
decreasing warp factor, while spin 1/2 and 3/2 fermionic fields are not
so in the string-like defect \cite{Oda}. 
Thus, it is natural to consider first the case of
a bulk scalar field. The case of a bulk vector field will be examined
in the next section. The spin 2 graviton was examined in detail
in Ref. \cite{Gherghetta} so we skip this case in this paper. 
From now on, for clarity we shall limit our attention to a 
$\it{local}$ string-like solution (\ref{20}) since the 
generalization to a $\it{global}$ solution (\ref{22}) is
straightforward. Of course, we have implicitly assumed that 
various bulk fields considered below make little contribution 
to the bulk energy so that the solution (\ref{20}) remains valid 
even in the presence of bulk fields.

Let us consider the action of a massless real scalar coupled
to gravity:
\begin{eqnarray}
S_\Phi = - \frac{1}{2} \int d^D x \sqrt{-g} g^{M N}
\partial_M \Phi \partial_N \Phi,
\label{24}
\end{eqnarray}
{}from which the equation of motion can be derived:
\begin{eqnarray}
\frac{1}{\sqrt{-g}} \partial_M (\sqrt{-g} g^{M N} \partial_N \Phi) = 0.
\label{25}
\end{eqnarray}
{}From now on we shall take $\hat{g}_{\mu\nu} = \eta_{\mu\nu}$
and define $P(r)=e^{-cr}$. 
In the background metric (\ref{20}), the equation of motion
(\ref{25}) reads
\begin{eqnarray}
P^{-1} \eta^{\mu\nu} \partial_\mu \partial_\nu \Phi
+ P^{-\frac{p+1}{2}} \partial_r (P^{\frac{p+1}{2}} 
\partial_r \Phi) + \frac{1}{R_0^2} P^{-1} \partial_\theta^2 \Phi
= 0.
\label{26}
\end{eqnarray}

Let the KK expansion of $\Phi$ be given by
\begin{eqnarray}
\Phi(x^M) = \sum_{n,l=0}^{\infty} \phi^{(n,l)}(x^\mu) 
\frac{\chi_n(r)} {\sqrt{R_0}} 
Y_l(\theta).
\label{27}
\end{eqnarray}
Here $Y_l(\theta)$ are in general the eigenfunction of the scalar
Laplacian $\Delta$ on a unit $(n-1)$-sphere with the eigenvalues
$l(l+n-2)$. Now we are taking account of a stringy defect with
codimension 2, i.e., $n$ is chosen to 2, so we have an equation
\begin{eqnarray}
\Delta Y_l(\theta) = l^2 Y_l(\theta),
\label{28}
\end{eqnarray}
with $l=0, 1, 2, \cdots$. And $Y_l(\theta)$ satisfy the following
orthonormality condition
\begin{eqnarray}
\int_0^{2\pi} d\theta Y_l(\theta) Y_{l'}(\theta)
= \delta_{ll'}.
\label{29}
\end{eqnarray}

Using the KK expansion (\ref{27}), the equation of motion
(\ref{26}) reduces to the well-known Klein-Gordon's equation
with the KK masses $m_n$:
\begin{eqnarray}
\left(\eta^{\mu\nu} \partial_\mu \partial_\nu -
m_n^2 \right) \phi^{(n,l)} = 0,
\label{30}
\end{eqnarray}
where we have required $\chi$ to satisfy the following differential 
equation:
\begin{eqnarray}
- \left( P^{-\frac{p-1}{2}} \partial_r P^{\frac{p+1}{2}} 
\partial_r - \frac{l^2}{R_0^2} \right) \chi_n = m_n^2 \chi_n.
\label{31}
\end{eqnarray}
Actually, it is easily shown that by means of Eqs. (\ref{27}), (\ref{28}),
(\ref{29}) and (\ref{31}) the starting action (\ref{24}) can be 
written as
\begin{eqnarray}
S_\Phi = - \frac{1}{2} \sum_{n,l=0}^{\infty} \int d^p x 
\left[ \eta^{\mu\nu} \partial_\mu \phi^{(n,l)}
\partial_\nu \phi^{(n,l)} + m_n^2 \phi^{(n,l)} \phi^{(n,l)}
\right],
\label{32}
\end{eqnarray}
where we have also used the orthonormality condition
\begin{eqnarray}
\int_{0}^{\infty} dr P^{\frac{p-1}{2}} 
\chi_n \chi_{n'} = \delta_{nn'}.
\label{33}
\end{eqnarray}

To analyse the scalar KK mass spectrum, it is necessary to
solve Eq. (\ref{31}) explicitly. Defining $M_n^2 = m_n^2 
- \frac{l^2}{R_0^2}$, $z_n = \frac{2}{c} M_n P^{-\frac{1}{2}}$
and $h_n = P^{\frac{p+1}{4}} \chi_n$, Eq. (\ref{31}) can be
written in the form
\begin{eqnarray}
\left[ \frac{d^2}{dz_n^2} + \frac{1}{z_n}\frac{d}{dz_n}
+ \left\{1 - \frac{1}{z_n^2} \left(\frac{p+1}{2} \right)^2 \right\} 
\right] h_n = 0,
\label{34}
\end{eqnarray}
which is nothing but the Bessel equation of order $\frac{p+1}{2}$.
Thus, the solutions are of the form
\begin{eqnarray}
\chi_n = \frac{1}{N_n} P^{-\frac{p+1}{4}} \left[ J_{\frac{p+1}{2}}
(z_n) + \alpha_n Y_{\frac{p+1}{2}}(z_n) \right],
\label{35}
\end{eqnarray}
where $N_n$ are the wavefunction normalization constants and
$\alpha_n$ are constant coefficients.
The differential operator in (\ref{31}) is self-adjoint provided
that one imposes the boundary conditions \cite{Gherghetta}
\begin{eqnarray}
\chi_n'(0) = \chi_n'(\infty) = 0.
\label{36}
\end{eqnarray}
These boundary conditions lead to the relations
\begin{eqnarray}
\alpha_n &=& - \frac{J_{\frac{p-1}{2}}(z_n(0))}
{Y_{\frac{p-1}{2}}(z_n(0))} \nn\\
&=& - \frac{J_{\frac{p-1}{2}}(z_n(\bar{r}))}
{Y_{\frac{p-1}{2}}(z_n(\bar{r}))},
\label{37}
\end{eqnarray}
where $\bar{r}$ indicates the infrared cutoff, which 
is taken to be an infinity at the end of calculations.
Incidentally, in deriving (\ref{37}) we have used the formula holding 
in the Bessel functions
\begin{eqnarray}
Z_\nu'(z) = Z_{\nu-1}(z) - \frac{\nu}{z}Z_\nu(z),
\label{38}
\end{eqnarray}
with $Z$ being $J$ or $Y$.
Now in the limit $M_n << c$, the KK masses can be derived from
the equation \cite{Pomarol}
\begin{eqnarray}
J_{\frac{p-1}{2}}(z_n(\bar{r})) = 0,
\label{39}
\end{eqnarray}
which gives us the approximate mass formula
\begin{eqnarray}
M_n = \frac{c}{2} (n + \frac{p}{4} - \frac{1}{2}) \pi
e^{-\frac{1}{2} c \bar{r}},
\label{40}
\end{eqnarray}
Moreover, the normalization constant $N_n$ takes the approximate
form in the limit $M_n << c$,
\begin{eqnarray}
N_n = \sqrt{c} \frac{z_n(\bar{r})}{2M_n} 
J_{\frac{p+1}{2}}(z_n(\bar{r})).
\label{41}
\end{eqnarray}

Note that in the limit $\bar{r} \rightarrow \infty$,
$M_n$ approach zero as in the graviton \cite{Gherghetta}, 
which is a characteristic feature of noncompact extra dimensions.
The KK masses of a scalar field are given by not $M_n$
but $m_n$, so it turns out that they approach $\frac{l^2}{R_0^2}$. 
Accordingly, only the $s$-wave ($l=0$) mode becomes massless 
on the string-like defect while the other modes are massive. Here it
is worth noticing that the massless $s$-wave mode is degenerate with
respect to the radial quantum number $n$ since the KK masses depend on
only $l$ in the limit $\bar{r} \rightarrow \infty$. Thus, we are
in danger of the existence of an infinite number of massless
modes on the defect, which seems to be against the phenomenology.
Luckily enough, however, as will be shown below, 
only a unique massless mode with
$n=0$ couples to fermion on the defect since the coupling constant
between the remaining massless modes with $n \ge 1$ and the defect
fermion vanishes in the infinite volume limit. It would be then natural to
identify this massless zero mode with $n=0$ as the Higgs field in our
world from the phenomenological viewpoint.

As is shown in Ref. \cite{Oda}, spin 1/2 fermion is localized on 
a defect with the exponentially rising warp factor, 
so it is necessary to invoke additional interactions 
except gravity in a model in order to localize spin 1/2 
fermion on our defect, which has the exponentially decreasing 
warp factor. In this letter, we simply consider fermion on
the string-like defect.
 
To see that only the massless zero mode with $n=0$ couples to 
fermion on the defect,
it is useful to examine the Yukawa coupling whose interaction
term is given by
\begin{eqnarray}
S_{\bar{\Psi}\Psi\Phi} = - g_\Phi \int d^D x \sqrt{-g}
\bar{\Psi} \Psi \Phi \delta(r).
\label{41-2}
\end{eqnarray}
The integration over the angular variable and the KK expansion
(\ref{27}) yield 
\begin{eqnarray}
S_{\bar{\Psi}\Psi\Phi} = - g_\Phi \sqrt{R_0} \int d^p x \bar{\Psi} \Psi
\sum_{n=0}^\infty \phi^{(n,0)}(x) \chi_n(0).
\label{41-3}
\end{eqnarray}
The wavefunction for the zero mode is a constant and from the
orthonormality condition (\ref{33}) we have the zero mode
\begin{eqnarray}
\chi_0 = \sqrt{\frac{c(p-1)}{2}}.
\label{41-4}
\end{eqnarray}
For the excited KK modes $\chi_n(0)$ with $n \ge 1$, it is easy to
evaluate $\chi_n(0)$ in the limit $M_n << c$
\begin{eqnarray}
\chi_n(0) = \frac{1}{N_n} J_{\frac{p+1}{2}}(\frac{2}{c} M_n) 
= \sqrt{c} P^{\frac{1}{4}}(\bar{r}),
\label{41-5}
\end{eqnarray}
where (\ref{41}) was used.
Moreover, defining the effective p-dimensional coupling constant
as $\tilde{g}_\Phi = g_\Phi \sqrt{\frac{c(p-1)}{2} R_0}$, 
the Yukawa interaction can be expressed as
\begin{eqnarray}
S_{\bar{\Psi}\Psi\Phi} = - \tilde{g}_\Phi \int d^p x \bar{\Psi}
\Psi \left[\phi^{(0,0)}(x) + \sqrt{\frac{2}{p-1}} P^{\frac{1}{4}}
(\bar{r}) \sum_{n=1}^\infty \phi^{(n,0)}(x) \right].
\label{41-6}
\end{eqnarray}
{}From this equation, it is obvious that the effective coupling of the excited
KK modes with $n \ge 1$ to the defect fermion vanishes in the 
limit $\bar{r} \rightarrow \infty$ owing to the presence of 
$P^{\frac{1}{4}} (\bar{r})$ in front of the second term. 
On the other hand, the massless zero mode with $n=0$ 
has a coupling constant of order one. Hence, only the massless
zero mode lives in the string-like defect.

\section{Kaluza-Klein decomposition of vector field}

Next we turn our attention to the case of vector field.
It was shown in the Randall-Sundrum model in $AdS_5$ space 
that spin 1 vector field is not localized neither
on a brane with positive tension nor on a brane with negative
tension so the Dvali-Shifman mechanism \cite{Dvali} must be
invoked for the vector field localization \cite{Bajc, Pomarol}.
On the other hand, we have shown that spin
1 vector field $\it{is}$ localized on a string-like
defect like spin 0 scalar and spin 2 graviton fields \cite{Oda}.
So we do not need to introduce additional mechanism for
the vector field localization in the case at hand.
The localization of vector field on the defect therefore allows
us to think of the bulk vector field.

Let us start with the action of the $U(1)$ vector field:
\begin{eqnarray}
S_A = - \frac{1}{4} \int d^D x \sqrt{-g} g^{M N} g^{R S}
F_{MR} F_{NS},
\label{42}
\end{eqnarray}
where $F_{MN} = \partial_M A_N - \partial_N A_M$ as usual.
(The extension to the case of non-Abelian gauge fields
is straightforward.)
{}From this action the equations of motion are given by
\begin{eqnarray}
\frac{1}{\sqrt{-g}} \partial_M (\sqrt{-g} g^{M N} g^{R S} F_{NS}) = 0.
\label{43}
\end{eqnarray}
With the background metric (\ref{20}) and the gauge conditions
$\partial_\mu A^\mu = A_{\theta} = 0$, these equations become
\begin{eqnarray}
\left(\eta^{\mu\nu} \partial_\mu \partial_\nu 
+ P^{\frac{3-p}{2}} \partial_r P^{\frac{p-1}{2}} \partial_r
+ \frac{1}{R_0^2} \partial_\theta^2 \right) A_\lambda
- P^{\frac{3-p}{2}} \partial_r P^{\frac{p-1}{2}} \partial_\lambda
A_r = 0,
\label{44}
\end{eqnarray}
\begin{eqnarray}
\left(\eta^{\mu\nu} \partial_\mu \partial_\nu 
+ \frac{1}{R_0^2} \partial_\theta^2 \right) A_r = 0,
\label{45}
\end{eqnarray}
\begin{eqnarray}
\partial_r \left( P^{\frac{p-1}{2}} \partial_\theta A_r
\right) = 0.
\label{46}
\end{eqnarray}

Let us take the following forms of the KK decomposition
for later convenience:
\begin{eqnarray}
A_\mu(x^M) &=& \sum_{n,l=0}^{\infty} A_\mu^{(n,l)}(x^\mu) 
\frac{f_n(r)}{\sqrt{R_0}} Y_l(\theta), \nn\\
A_r(x^M) &=& \sum_{l=0}^{\infty} A_r^{(l)}(x^\mu) 
\frac{g(r)}{\sqrt{R_0}} Y_l(\theta).
\label{47}
\end{eqnarray}
Then, from Eq. (\ref{45}) we have
\begin{eqnarray}
\left(\eta^{\mu\nu} \partial_\mu \partial_\nu 
- \frac{l^2}{R_0^2} \right) A_r^{(l)}(x^\mu) = 0.
\label{48}
\end{eqnarray}
In addition, Eq. (\ref{46}) leads to a general solution for
$g(r)$.
\begin{eqnarray}
g(r) = \alpha P^{-\frac{p-1}{2}},
\label{49}
\end{eqnarray}
with $\alpha$ being an integration constant.
Finally, using (\ref{49}), Eq. (\ref{44}) reduces to the form
\begin{eqnarray}
\left(\eta^{\mu\nu} \partial_\mu \partial_\nu 
- m_n^2 \right) A_\lambda^{(n,l)}(x^\mu) = 0.
\label{50}
\end{eqnarray}
Here we have required $f_n(r)$ to satisfy the differential
equation
\begin{eqnarray}
- \left( P^{-\frac{p-3}{2}} \partial_r P^{\frac{p-1}{2}} 
\partial_r - \frac{l^2}{R_0^2} \right) f_n(r) = m_n^2 f_n(r).
\label{51}
\end{eqnarray}

As in the case of a scalar field, let us substitute the KK 
expansion (\ref{47}) and the solution (\ref{49}) into the starting 
action (\ref{42}), whose result is given by
\begin{eqnarray}
S_A = \int d^p x \sum_{n,l=0}^{\infty} \left[-\frac{1}{4} 
\eta^{\mu\nu} \eta^{\lambda\rho} F_{\mu\lambda}^{(n,l)} 
F_{\nu\rho}^{(n,l)} - \frac{1}{2} m_n^2 \eta^{\mu\nu} A_\mu^{(n,l)} 
A_{\nu}^{(n,l)} \right] \nn\\
- \frac{\alpha^2}{c(p-1)} [P(\bar{r})^{-\frac{p-1}{2}} - 1]
\times \int d^p x \sum_{l=0}^{\infty} \left[\eta^{\mu\nu} 
\partial_\mu A_r^{(l)} \partial_\nu A_r^{(l)} + \frac{l^2}{R_0^2}
A_r^{(l)} A_r^{(l)} \right],
\label{52}
\end{eqnarray}
where we have used (\ref{29}) and the orthonormality condition
for $f_n(r)$
\begin{eqnarray}
\int_{0}^{\infty} dr P^{\frac{p-3}{2}} 
f_n f_{n'} = \delta_{nn'}.
\label{53}
\end{eqnarray}
Note that the coefficient in front of the action of the scalar
field $A_r$ becomes divergent in the limit $\bar{r} \rightarrow
\infty$, but this divergence can be absorbed in the redefinition of
the field $A_r^{(l)}$. That is, by performing the field-redefinition
$A_r^{(l)} \rightarrow \alpha \sqrt{\frac{2}{c(p-1)}} 
\sqrt{P(\bar{r})^{-\frac{p-1}{2}} - 1} A_r^{(l)}$, we arrive at the 
expression
\begin{eqnarray}
S_A = \int d^p x \sum_{n,l=0}^{\infty} \left[-\frac{1}{4} 
\eta^{\mu\nu} \eta^{\lambda\rho} F_{\mu\lambda}^{(n,l)} 
F_{\nu\rho}^{(n,l)} - \frac{1}{2} m_n^2 \eta^{\mu\nu} A_\mu^{(n,l)} 
A_{\nu}^{(n,l)} \right] \nn\\
- \frac{1}{2} \int d^p x \sum_{l=0}^{\infty} \left[\eta^{\mu\nu} 
\partial_\mu A_r^{(l)} \partial_\nu A_r^{(l)} + \frac{l^2}{R_0^2}
A_r^{(l)} A_r^{(l)} \right].
\label{52-2}
\end{eqnarray}
Here notice that the second integral describes that the 'gauge-scalar'
has the same structure as the action of scalar field (33). Thus,
following the similar argument to the case of a scalar field, it is
straightforward to show that only the massless zero mode of the 
'gauge-scalar' couples to fermion on the defect. Therefore, 
in what follows, we shall consider the $p$-dimensional gauge field $A_\mu$.

To examine the KK spectrum, we can follow a similar path of argument
to that of a scalar field in Section 3.
This time, by defining $h_n = P^{\frac{p-1}{4}} f_n$, 
Eq. (\ref{51}) can be written as
\begin{eqnarray}
\left[ \frac{d^2}{dz_n^2} + \frac{1}{z_n}\frac{d}{dz_n}
+ \left\{1 - \frac{1}{z_n^2} \left(\frac{p-1}{2} \right)^2 \right\} 
\right] h_n = 0,
\label{54}
\end{eqnarray}
whose solution is also expressed in terms of 
the Bessel functions of order $\frac{p-1}{2}$
\begin{eqnarray}
f_n (z_n) = \frac{1}{N_n} P^{-\frac{p-1}{4}} \left[ J_{\frac{p-1}{2}}
(z_n) + \alpha_n Y_{\frac{p-1}{2}}(z_n) \right],
\label{55}
\end{eqnarray}
where $N_n$ are new wavefunction normalization constants and
$\alpha_n$ are new constant coefficients.
The same boundary conditions (\ref{36}) for $f_n(r)$ lead to the relations
\begin{eqnarray}
\alpha_n &=& - \frac{J_{\frac{p-3}{2}}(z_n(0))}
{Y_{\frac{p-3}{2}}(z_n(0))} \nn\\
&=& - \frac{J_{\frac{p-3}{2}}(z_n(\bar{r}))}
{Y_{\frac{p-3}{2}}(z_n(\bar{r}))}.
\label{56}
\end{eqnarray}

In the limit $M_n << c$, the KK masses can be derived from
the equation
\begin{eqnarray}
J_{\frac{p-3}{2}}(z_n(\bar{r})) = 0,
\label{57}
\end{eqnarray}
which gives us the approximate mass formula
\begin{eqnarray}
M_n = \frac{c}{2} (n + \frac{p}{4} - 1) \pi
e^{-\frac{1}{2} c \bar{r}},
\label{58}
\end{eqnarray}
And the normalization constant $N_n$ takes the approximate
form 
\begin{eqnarray}
N_n = \sqrt{c} \frac{z_n(\bar{r})}{2M_n} 
J_{\frac{p-1}{2}}(z_n(\bar{r})).
\label{59}
\end{eqnarray}

Note that in the limit $\bar{r} \rightarrow \infty$,
the KK masses of vector field are given by $\frac{l^2}{R_0^2}$
like the scalar case. 
Hence as expected, only the $s$-wave ($l=0$) becomes massless on 
the p-brane defect while the other modes are massive. This time,
compared to the scalar case, it is more important to show that 
only one massless mode with $n=0$ lives in the defect since 
such a massless zero mode would be regarded as the unique 'photon' 
on the defect.

We are now ready to consider the coupling of the gauge KK modes
to spin 1/2 fermion on the p-brane defect. 
The fermion kinetic and gauge interaction terms are given by
\begin{eqnarray}
S_\Psi = \int d^D x \sqrt{-g} \bar{\Psi} i \Gamma^M 
(\partial_\mu + i g_A A_\mu) \Psi \delta_M^\mu \delta(r),
\label{60}
\end{eqnarray}
where the curved gamma matrices $\Gamma^\mu$ and the flat gamma ones
$\gamma^\mu$ are related through the relations 
$\Gamma^\mu = P^{-\frac{1}{2}} \gamma^\mu$. Upon integrating
over $\theta$ and using the KK expansion (\ref{47}), we obtain for
the gauge-fermion interaction term
\begin{eqnarray}
S_{\bar{\Psi}\Psi A} = - g_A \sqrt{R_0} \int d^p x \bar{\Psi} \gamma^\mu 
\sum_{n=0}^\infty A_\mu^{(n,0)}(x) f_n(0) \Psi.
\label{61}
\end{eqnarray}

The wavefunction for the zero mode is again a constant and the
orthonormality condition (\ref{53}) gives us the zero mode
\begin{eqnarray}
f_0 = \sqrt{\frac{c(p-3)}{2}}.
\label{62}
\end{eqnarray}
Recall that this zero mode is localized on the string-like
defect and is identified with the usual 'photon' of the 
p-dimensional Minkowski space-time \cite{Oda} while it is not 
localized on the domain wall \cite{Pomarol, Bajc}.
For the excited KK modes $f_n(0)$ with $n \ge 1$, it is easy to
evaluate $f_n(0)$ in the limit $M_n << c$
\begin{eqnarray}
f_n(0) = \frac{1}{N_n} J_{\frac{p-1}{2}}(\frac{2}{c} M_n) 
= \sqrt{c} P^{\frac{1}{4}}(\bar{r}),
\label{63}
\end{eqnarray}
where (\ref{59}) was used.
Then, defining the effective p-dimensional $U(1)$ coupling constant
as $\tilde{g}_A = g_A \sqrt{\frac{c(p-3)}{2} R_0}$, the interaction term
reads
\begin{eqnarray}
S_\Psi^{int} = - \tilde{g}_A \int d^p x \bar{\Psi} \gamma^\mu 
\left[A_\mu^{(0,0)}(x) + \sqrt{\frac{2}{p-3}} P^{\frac{1}{4}}
(\bar{r}) \sum_{n=1}^\infty A_\mu^{(n,0)}(x) \right] \Psi.
\label{64}
\end{eqnarray}

{}From this equation, it is obvious that the coupling of the excited
KK modes to the defect fermion vanishes in the limit $\bar{r}
\rightarrow \infty$ owing to the presence of $P^{\frac{1}{4}}
(\bar{r})$ as in the scalar field. On the other hand, the
massless zero mode has a coupling constant of order one as desired.
Thus, this model is consistent with gauge fields existing in
the bulk, which should be contrasted with the Randall-Sundrum
model \cite{Rizzo, Pomarol}. Of course, further studies are
necessary to assure the consistency of the model at hand at the
quantum level.

\section{Discussions}

In this paper we have explored the possibility of placing spin 0
scalar field and spin 1 vector gauge field in the bulk in the 
string-like defect model in detail. We have derived the scalar and the gauge
field KK spectra from examination of the action of the theory and
also analyzing the equations of motion. 

We then computed the scalar-fermion and the gauge-fermion interactions 
on the string-like defect and found that the excited KK states with
respect to the radial quantum number do not couple to fermion on
the defect in the infinite cutoff limit, whereas the massless zero modes,
which are nothing but the 'Higgs' particle and the usual 'photon' 
of the Minkowski space-time in the cases of scalar and gauge boson,
respectively, couple to fermion with order unit. 
Since it has been already shown that spin 2 graviton is localized 
on the defect \cite{Oda} and yields the desired Newton's law 
with tiny correction terms on the defect \cite{Gherghetta},
the model which we consider equips with desirable physical properties.
Of course, to make the model at hand more realistic we need additional
interactions except gravity for localizing fermions on the defect, but we
wish 
to insist that a supergravity model corresponding to the present 
model would resolve this problem in a natural way 
even if we could localize fermionic fields by introducing additional 
interaction terms between bosons and fermions by hand. 

Let us restrict the following argument to the case of $p=4$ and $D=6$.
In this case, it is of interest to imagine that 10D superstring theory 
might be compactified on Calabi-Yau two-fold, i.e., K3 by the conventional 
Kaluza-Klein mechanism, yielding 6D theory and then the 6D theory 
is compactified to our four dimensional space-time according to 
the alternative compactification scenario discussed in this paper. 
Here it is worthwhile to mention that the 6D local field theory is a
very interesting field theory. (Correspondingly, 6D supergravity theory 
possesses a richer structure than 5D supergravity theory in many respects.) 
For instance, the 6D local field theory is a free 
theory with trivial cubic scalar self-interaction with unrenormalizable
Einstein-Hilbert and Yang-Mills actions, so it is expected that 'little' 
superstring theory may play an essential role. Furthermore, our model 
in 6 space-time dimensions has a physical setting where our world is 
a 3-brane embedded in 6D space-time with non-factorizable warped geometry. 
Interestingly enough, the metric of two internal dimensions is conformally 
flat, so the powerful (Euclidean) conformal field theory technique can 
be naturally applied to our model.   
   
Finally let us comment on a supersymmetric realization
of the present model. A lot of attention has been recently devoted to
the construction of a supersymmetric Randall-Sundrum model 
\cite{Kallosh1, Behrndt, Bagger, de Alwis, Maldacena, 
Kallosh2, Duff} and advocated some no-go theorems. Then it is 
of interest to ask whether there are supersymmetric relations between 
our 6D model and 5D Randall-Sundrum model. Through a simple KK dimensional
reduction, it seems that the background metric in our model reduces to the
one in the Randall-Sundrum model. At the same time, N=2, 6D supergravity
would reduce to N=2, 5D supergravity. Thus, if we cannot construct
a supersymmetric version of the Randall-Sundrum model, it might be also
difficult to construct a supersymmetric model of our 6D theory.
But there recently appeared an interesting construction of a supersymmetric
Randall-Sundrum model \cite{Duff}. The corresponding construction of
our 6D model is now in an active investigation so we hope to report this
construction in near future.

\vs 1
\begin{flushleft}
{\bf Acknowledgement}
\end{flushleft}
We are indebted to M. Tonin for valuable discussions and 
continuous encouragement. 
We wish to thank Dipartimento Di Fisica, "Galileo Galilei", 
Universita Degli Studi Di Padova, for a kind hospitality, where 
most of parts of this work have been done.

\vs 1

\end{document}